\documentclass{iau}

\usepackage{amsmath}
\usepackage{graphicx}
\usepackage{orcidlink}
\usepackage{xspace}
\usepackage{aas_macros}
\usepackage{adjmulticol}
\usepackage[font=small]{caption}
\usepackage[final]{microtype} 

\newcommand{\kms}{\,km\,s$^{-1}$\xspace} 
\newcommand{\Msun}{$\,{\rm M}_\odot$\xspace}  
\newcommand{\Rsun}{$\,{\rm R}_\odot$\xspace}  
\newcommand{\Zsun}{$\,{\rm Z}_\odot$\xspace}  
\newcommand{\logg}{$\log g$\xspace} %
\newcommand{\drv}{$\Delta$RV\xspace} %

\begin{document}

\lefttitle{J.~I. Villaseñor}
\righttitle{The binary landscape of massive stars at low Z: Insights from the BLOeM Campaign}

\jnlPage{1}{7}
\jnlDoiYr{2021}
\doival{10.1017/xxxxx}
\volno{402}

\aopheadtitle{Proceedings IAU Symposium}
\editors{A. Wofford, N. St-Louis, M. García \& S. Simón-Díaz, eds.}

\title{The binary landscape of massive stars at low Z: Insights from the BLOeM Campaign}

\author{
    J.~I. Villaseñor$^{1}$\orcidlink{0000-0002-7984-1675}, 
    H. Sana$^{2}$\orcidlink{0000-0001-6656-4130},
    J. Bodensteiner$^{3}$\orcidlink{0000-0002-9552-7010},
    N. Britavskiy$^{4}$\orcidlink{0000-0003-3996-0175},
    L.~R. Patrick$^{5}$\orcidlink{0000-0002-9015-0269},
    T. Shenar$^{6}$\orcidlink{0000-0003-0642-8107} and the BLOeM Collaboration}
\affiliation{
    $^{1}$Max-Planck-Institut für Astronomie, Königstuhl 17, D-69117 Heidelberg, Germany \\
    $^{2}$Institute of Astronomy, KU Leuven, Celestijnenlaan 200D, 3001 Leuven, Belgium \\
    $^{3}$Anton Pannekoek Institute for Astronomy, University of Amsterdam, Science Park 904, 1098 XH Amsterdam, The Netherlands \\
    $^{4}$Royal Observatory of Belgium, Avenue Circulaire/Ringlaan 3, B-1180 Brussels, Belgium \\
    $^{5}$Centro de Astrobiolog\'{\i}a (CSIC-INTA), Ctra.\ Torrej\'on a Ajalvir km 4, 28850 Torrej\'on de Ardoz, Spain \\
    $^{6}$The School of Physics and Astronomy, Tel Aviv University, Tel Aviv 6997801, Israel
}

\begin{abstract}
We present an overview of our recent results from the BLOeM campaign in the Small Magellanic Cloud ($Z=0.2$\Zsun). Using nine-epoch VLT/FLAMES spectroscopy, we investigated the multiplicity of 929 massive stars. Our findings reveal contrasting binary properties across evolutionary stages: O-type stars show an intrinsic close-binary fraction of $70\%$, and early B-type dwarfs/giants reach $\sim80\%$, exceeding higher-metallicity samples. In contrast, B0--B3 supergiants drop to $\sim40\%$, and A--F supergiants to $\sim8\%$; intrinsic variability likely inflates the latter, so the true multiplicity may be lower. OBe stars display distinct binary properties consistent with a post-interaction origin. These results have profound implications for massive-star evolution at low metallicity, including the production of exotic transients, gravitational-wave progenitors, and ionising radiation in the early Universe.
\end{abstract}


\maketitle

\section{Introduction}

Over the past 15 years, binarity has emerged as central to massive-star evolution. Spectroscopic surveys of young Galactic clusters report high intrinsic close binary fractions: $f_{\rm bin}=69\pm9\%$ for O stars \citep{sana+12}, $55\%$ for OB stars in Cyg~OB2 \citep{kobulnicky+14}, and $53\pm8\%$ across B0--B9 in NGC~6231 \citep{banyard+22}. Given these rates and the preference for short periods, most systems with $P\lesssim3000$\,d are expected to interact, producing mergers and stripped stars \citep{sana+12,demink+14}. In the Large Magellanic Cloud (LMC), the VLT-FLAMES Tarantula Survey \citep[VFTS,][]{evans+11} extended this work to lower metallicity, with $Z_{\rm LMC}\approx0.5$\Zsun. Among 768 stars earlier than B3, the intrinsic O-star binary fraction was $51\pm4\%$ \citep{sana+13}, later suggested to be $\sim60\%$ \citep{almeida+17}. For B-type stars with masses of 8--15\Msun, \cite{dunstall+15} found $58\pm11\%$ for dwarfs (\logg${>}3.3$) and $f_{\rm obs}=23\pm6\%$ for supergiants. Because the ${\sim}6$-epoch cadence was insufficient to derive orbital solutions, two follow-ups, the Tarantula Massive Binaries Monitoring \citep[TMBM;][]{almeida+17} for O stars and the B-type Binaries Characterisation programme \citep[BBC;][]{villasenor+21}, added 32 and 29 FLAMES/GIRAFFE epochs, deriving orbital solutions and constraining the period, eccentricity, and mass-ratio distributions. Together with Galactic results, these studies found that binary fractions and orbital period distributions in the LMC and the Milky Way (MW) were remarkably similar, with no strong differences between O and B stars. However, the question of whether even lower metallicities, like those representative of the early Universe, affect the multiplicity properties of massive stars remained.

\begin{figure}[t]
    \centering
    \includegraphics[width=0.75\linewidth]{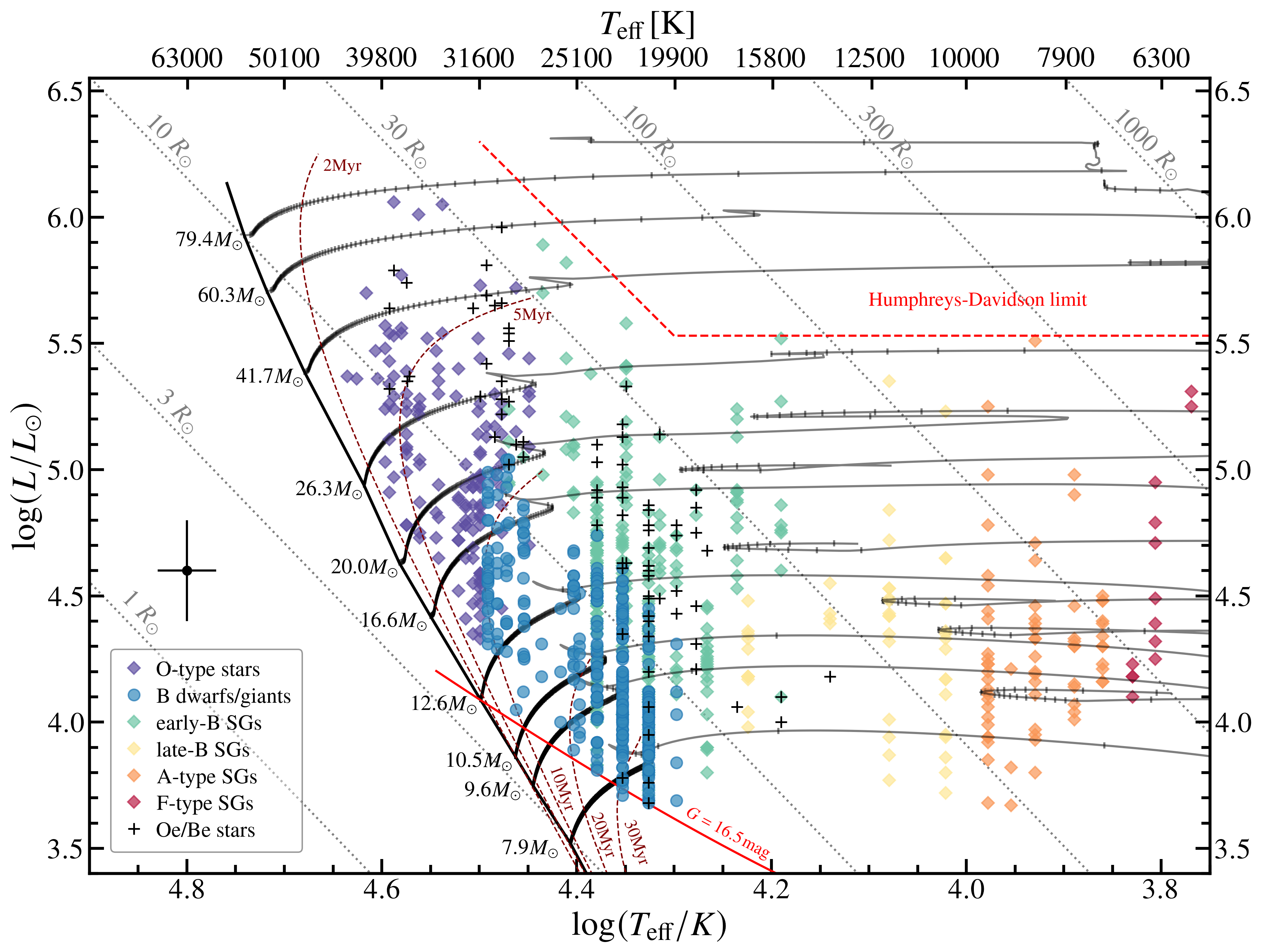}
    \caption{Hertzsprung-Russell diagram of the BLOeM sample in the SMC, colour-coded by subsample (legend). From \cite{villasenor+25}.}
    \label{fig:hrd}
\end{figure}

\section{The BLOeM campaign}

At 62\,kpc \citep{graczyk+20} and $Z_{\rm SMC}=0.2$\Zsun, the Small Magellanic Cloud (SMC) serves as an excellent local analogue of high-$z$ star-forming galaxies \citep{nakajima+23}. However, spectroscopic surveys targeting a sufficiently large sample of OB-type stars with enough cadence and baseline to determine their multiplicity properties were severely lacking. 

The Binarity at LOw Metallicity campaign \citep[BLOeM;][]{shenar+24} was designed to provide a homogeneous, multi-epoch census: 25 VLT/FLAMES-GIRAFFE epochs for 929 OBAF-type stars in eight SMC fields, with sampling and baseline comparable to TMBM and BBC in 30~Doradus, enabling robust radial-velocity (RV) variability detection and orbital solutions at low $Z$.

\section{Binary fraction at SMC metallicity}

Given the heterogeneity of the sample (O dwarfs to F supergiants), the sample was split in subsamples by spectral types and luminosity classes from \cite{shenar+24} as shown in Fig.~\ref{fig:hrd}: O-type stars \citep{sana+25}; B-type dwarfs and giants \citep[B0--B3 V--III;][]{villasenor+25}; emission-line stars \citep[Oe/Be;][]{bodensteiner+25}; early B bright giants and supergiants \citep[B0--B3 II--I;][]{britavskiy+25}; and late supergiants \citep[B4--F II--I;][]{patrick+25}. These studies focused on the multiplicity fraction using the first nine epochs (October--December 2023), with baselines from ${\sim}30$\,d (Field~6) to ${\sim}66$\,d (Field~1).

\vspace{-5mm}
\subsection{O-type stars}

\cite{sana+25} analysed 139 O-type stars, deriving RVs from cross-correlation and line-profile fitting. A system is classified as binary if (i) the amplitude of the RV variation (\drv) is ${>}20$\kms and (ii) any epoch pair satisfies \drv$/\sigma_{\Delta\rm RV}>4$. These criteria have been widely used in massive-star multiplicity studies \citep[e.g.][]{sana+13, bodensteiner+21,banyard+22} and in other BLOeM works, ensuring comparability. A total of 62 stars meet both criteria, translating to an observed binary fraction of $f_{\rm obs}=45\pm4\%$.

To compare intrinsic fractions across samples with different cadences, baselines, and magnitudes, \citep{sana+25} modelled the survey's detectability via $10^4$ Monte Carlo observing campaigns that replicate BLOeM’s cadence and RV uncertainties. Binaries are drawn from power-law distributions in $\log P$ (index $\pi$), mass ratio (approximately flat), and eccentricity ($\eta\simeq-0.5$) over $P\simeq1$--$3000$\,d and $M_1\simeq15$--$60$\Msun. Applying the same criteria to the mock samples gives detection probabilities $>0.9$ for $P\lesssim3$ months, dropping steeply beyond $\sim100$\,d (their Fig.~2). They then conduct a grid search over $(f_{\rm bin},\pi)$, selecting models that jointly match the observed number of binaries (binomial likelihood for $N_{\mathrm{bin}}$) and the distribution of shortest significant time lapses $\delta t$ (two-sided Kuiper statistic). The preferred model yields $f_{\rm bin}=0.70^{+0.11}_{-0.06}$ and a near-flat period index $\pi=+0.10^{+0.20}_{-0.15}$ (their Fig.~3). Accounting for line blending in near-equal components would raise $f_{\rm bin}$ by $\sim5\%$, so these are conservative lower limits.

\vspace{-5mm}
\subsection{B-type dwarfs and giants}

The non-supergiant B sample is the largest, with 309 B0--B2.5 V--III stars; the later subtypes (B1--B2.5) are dominated by giants because of the BLOeM magnitude limit ($G<16.5$\,mag). \cite{villasenor+25} measured RVs using pure line-profile fitting: least-squares for SB1 systems and a hierarchical Bayesian two-component model for SB2s that fits all lines and epochs simultaneously \citep[as in][]{sana+13}, with parameters inferred via Hamiltonian Monte Carlo (NUTS). Applying the same binary criteria as \cite{sana+25} yields $f_{\rm obs}=50\pm3\%$, consistent with the O-star value within uncertainties, despite lower signal-to-noise ratio (SNR) and stronger line-blending biases for B stars \citep[lower-mass primaries; see][]{sana+25}.

\begin{figure}[t]
\centering
\begin{minipage}{0.48\linewidth}
    \centering
    \includegraphics[width=\linewidth]{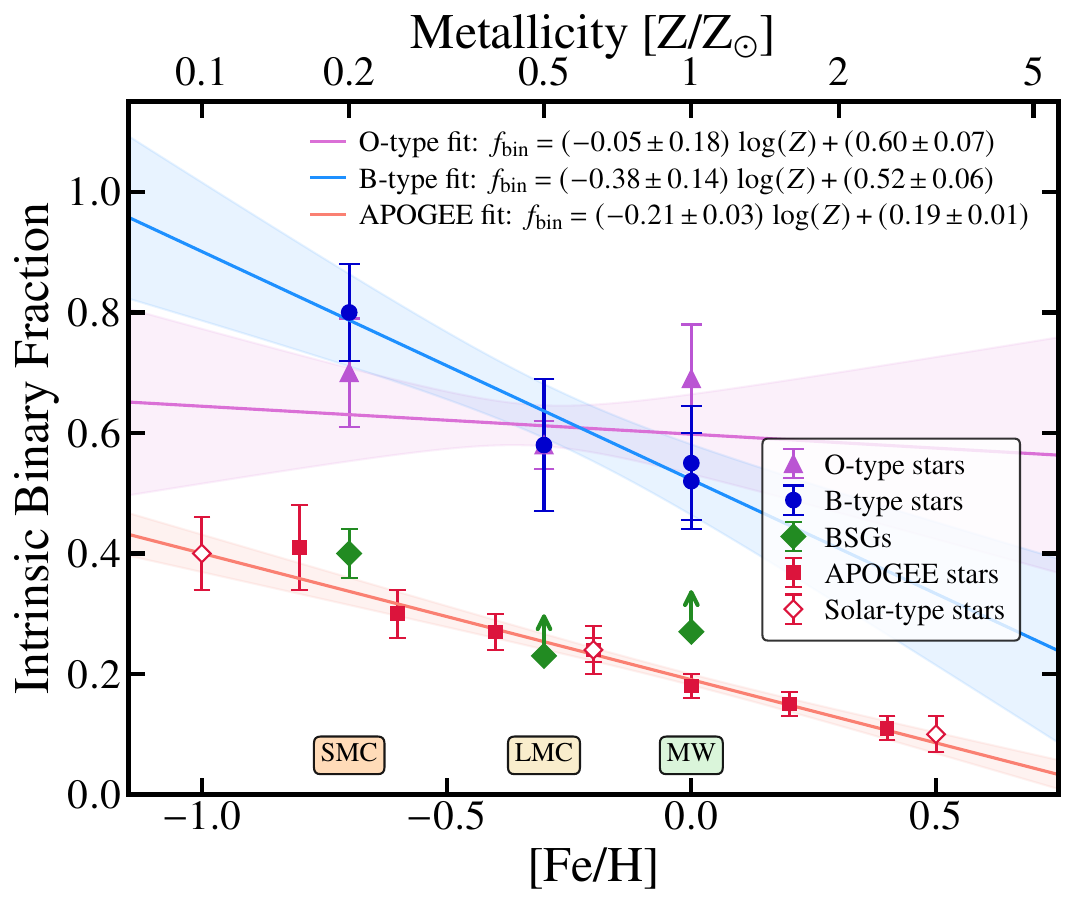}
    \caption{Intrinsic close-binary fraction versus metallicity for O and B stars in the MW, LMC, and SMC, including BSGs and the APOGEE solar-type trend for comparison. Adapted from \cite{villasenor+25} and \cite{sana+25}.}
    \label{fig:fbin_Z}
\end{minipage}\hfill
\begin{minipage}{0.48\linewidth}
    \centering
    \includegraphics[width=\linewidth]{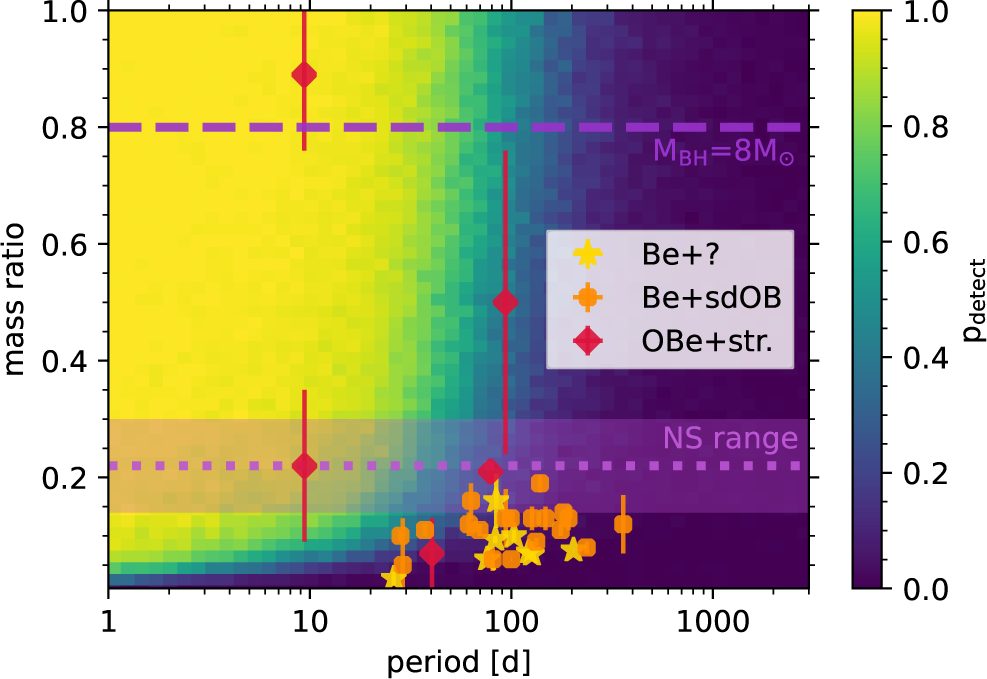}
    \caption{Binary detection probability $p_{\rm detect}$ as a function of period and mass ratio for a 10\Msun Be star. Overplotted are literature systems, see details in \cite{bodensteiner+25}.}
    \label{fig:obe}
\end{minipage}
\end{figure}

To infer the intrinsic binary fraction, \citep{villasenor+25} first applied the same framework as for the O stars, obtaining $f_{\rm bin}=80\pm8\%$. Because pulsations at low metallicity and the exact $\Delta\rm RV$ threshold for binarity could bias a criteria-based approach, they also fitted the full $\Delta\rm RV$ distribution with a Markov Chain Monte Carlo (MCMC) procedure, without imposing hard binary cuts, yielding $f_{\rm bin}=79\pm5\%$. Finally, including the orbital-period distribution in the forward modelling, parameterised as a power law in $\log_{10} P$ with index $\pi$, gave $f_{\rm bin}=85^{+7}_{-9}\%$ and $\pi=0.16\pm0.15$ (their Fig.~18), consistent with the O-star result of $\pi=0.10^{+0.20}_{-0.15}$ \citep[Fig.~3 in][]{sana+25}. All three estimates agree within $1\sigma$, indicating a robust intrinsic binary fraction near 80--85\% and no significant differences between the distribution of orbital periods of O- and early B-type stars.

Figure~\ref{fig:fbin_Z} compares intrinsic binary fractions for O and B stars in the MW, LMC, and SMC, including the anti-correlation reported for solar-type stars \citep{badenes+18,moe+19}. For O stars, \cite{sana+25} measured a slope $m=-0.05\pm0.18$, consistent with no trend, although not excluding the low-mass behaviour. For early B stars, \cite{villasenor+25} found an anti-correlation consistent with that of the APOGEE stars, $2.6\sigma$ from $m=0$, hinting that the B-star binary fraction increases toward low metallicity. Whether this reflects metallicity itself or environmental differences (e.g. field versus clusters) remains to be determined.

\begin{figure}[t]
\centering
\begin{minipage}{0.48\linewidth}
    \centering
    \includegraphics[width=\linewidth]{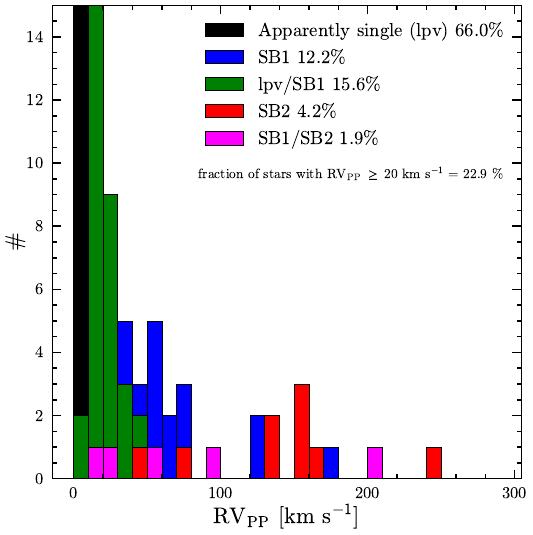}
    \caption{\drv distribution for early BSGs by spectroscopic status. Y axis limited to 15; first two bins: 150 and 24. From \cite{britavskiy+25}.}
    \label{fig:bsg_drv}
\end{minipage}\hfill
\begin{minipage}{0.48\linewidth}
    \centering
    \includegraphics[width=\linewidth]{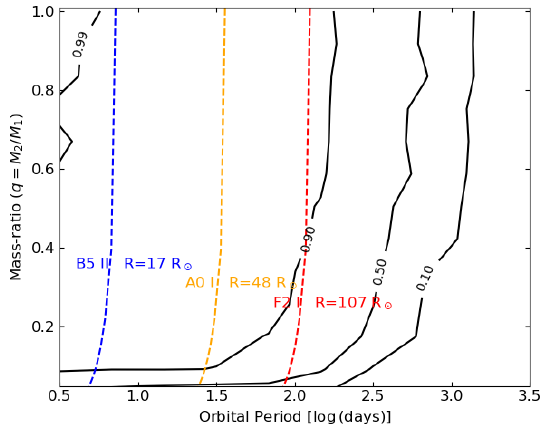}
    \caption{Orbital period versus mass ratio parameter space for simulated samples. Contours show detection probability ($p_{\rm detect}=99,90,50,10\%$). Dashed lines mark the minimum allowed period for three representative cases. From \cite{patrick+25}.}
    \label{fig:baf_fbin_dRV}
\end{minipage}
\end{figure}

\vspace{-5mm}
\subsection{The OBe stars}

Whether OBe stars are primarily products of binary interaction or can reach near-critical rotation on their own remains debated. \cite{bodensteiner+25} examined the BLOeM Oe/Be sample to provide further constraints. They analysed 18 Oe and 62 Be stars, measured RVs via cross-correlation, and found a deficit of large-amplitude RV variables; most detected binaries show \drv in the 20--50\kms range (their Fig.~3). Using the same binary criteria as above, they obtained an observed fraction of $f_{\rm obs}=18\pm4\%$, well below the O- and B-type values (45--50\%). No bias correction was applied due to the unconstrained distribution of orbital properties of this sample, and their uncertain evolutionary stage.

To assess detection capabilities, simulations over wide ranges of period and mass ratio were run using BLOeM’s cadence and uncertainties. The detection probability is near unity for $P<100$\,d across most of parameter space (Fig.~\ref{fig:obe}), indicating that main-sequence companions similar to those of O and B stars would have been found. The absence of such systems implies a different companion mass and period distribution, consistent with many Oe/Be stars being post-interaction products.

\vspace{-5mm}
\subsection{Early B supergiants and bright giants}

The early B-type SG (BSG) sample comprises 262 B0--B3 II--I stars. These bright targets reach SNR 70--100. \cite{britavskiy+25} measured RVs via cross-correlation and applied the same binary criteria as for the other samples. A large fraction shows line-profile variability (LPV; Fig.~\ref{fig:bsg_drv}), which complicates identifying shifts from orbital motion. Considering systems classified as LPV/SB1 (single-lined spectroscopic binaries) that satisfy both criteria, \cite{britavskiy+25} obtained $f_{\rm obs}=23\pm3\%$, rising to $f_{\rm bin}=40\pm4\%$ after bias corrections. The low observed fraction is expected given the evolutionary status; with atmospheres expanding beyond ${\sim}10$\Rsun, most systems with $M_{\rm ini}{>}8$\Msun, $q{>}0.5$, and $P{<}5$\,d are expected to have undergone Roche-lobe overflow. While the global fraction resembles that of the OBe stars, there are 11 BSGs with \drv$>100$\kms (Fig.~\ref{fig:bsg_drv}) that may be pre-interaction binaries or bloated stripped stars \citep[e.g.][]{villasenor+23}. For comparison, only two such systems are present in the OBe sample, with one likely interacting.

The intrinsic BSG fraction is compared with LMC \citep{dunstall+15} and MW \citep{deburgos+25} values in Fig.~\ref{fig:fbin_Z} (green diamonds). These higher-metallicity studies are not corrected for biases and thus give lower limits, precluding a direct comparison. Metallicity-dependent winds and pulsations also prevent a uniform RV-variability criterion, and varying cadence/baseline further hinder comparing observed fractions across surveys.

\vspace{-4mm}
\subsection{The BAF supergiants}

BLOeM also observed 128 cooler SGs spanning the B5--F5 spectral range. Owing to the larger number of narrow lines and SNR$\gtrsim70$, \cite{patrick+25} obtained precise RVs via cross-correlation. They found a dearth of large-amplitude RV variables: using the 20\kms threshold adopted for other samples would yield a binary fraction consistent with zero (their Fig.~4). In light of the RV precision and the level of intrinsic variability, they adopted a 5\kms threshold, finding $f_{\rm obs}^{\rm B}=25\pm6\%$ for late BSGs and $f_{\rm obs}^{\rm AF}=5\pm2\%$ for A/F supergiants. For late BSGs, the fraction agrees with the early BSGs; however, the lower RV threshold increases potential contamination from intrinsic variability, so the true fraction could be lower.

To infer intrinsic fractions, \cite{patrick+25} followed the \cite{sana+25} bias-correction method. Accounting for the larger radii of BAF stars, they obtained $f_{\rm bin}^{\rm B}=18^{+20}_{-16}\%$ for late BSGs and $f_{\rm bin}^{\rm AF}=8^{+9}_{-7}\%$ for A/F SGs. Notably, there is a lack of stars with \drv$>14$\kms. Given that systems with $R>17$\Rsun (up to a few dozen \Rsun), $q>0.1$, and $P_{\rm orb}\gtrsim6$\,d would be detected with $>90\%$ probability (Fig.~\ref{fig:baf_fbin_dRV}), this absence suggests the companions to BAF supergiants do not follow the multiplicity properties of the B-dwarf/giant population from \citealt{villasenor+25} \citep[see also Fig.~9 in][]{patrick+25}. Adding to the puzzle, the intrinsic binary fraction of red supergiants (RSGs) in the SMC is $f_{\rm bin}^{\rm RSG}=18.8\pm1.5\%$ \citep{patrick+22}, which is difficult to reconcile with a post-RSG population showing an even lower fraction.

\section{Conclusions}

BLOeM has delivered first multiplicity results at SMC metallicity for a large sample spanning young O-type dwarfs to F-type SGs. We confirm the high binary fraction of O-type stars and the dominance of post-interaction products in the OBe population, and we raise new questions: Why do B-type dwarfs and giants show such a high binary fraction at low metallicity? What is the origin of the BAF SGs? How strongly do pulsations affect O- and B-type dwarfs and SGs at low metallicity, and what is a reasonable RV-variability threshold for binarity?

The full set of 25 epochs will refine these results by yielding orbital solutions and orbital-property distributions. Further progress will require confronting star-formation scenarios and environmental effects (clusters versus field) on the binary fraction, together with multi-wavelength observations to search for hot companions and degenerate objects and to detect and characterise pulsations.

\begin{acknowledgements}
JIV acknowledges support from the European Research Council for the ERC Advanced Grant 101054731.
\end{acknowledgements}

\bibliographystyle{iaulike-1auth}
\begin{adjmulticols}{2}{-5mm}{-8mm}
\bibliography{JVbiblio} 
\end{adjmulticols}

\end{document}